\documentstyle[amssymb,12pt]{article}

\input{tcilatex}
\begin{document}

\title{{\bf {Type IIB String Backgrounds on Parallelizable PP-Waves and Conformal
Liouville Theory}.}}
\author{M. HSSAINI$^{1}$ and M.B.SEDRA$^{2}$ \thanks{
Corresponding author: sedra@ictp.trieste.it} \\
$^{1,\,2}${\small \ International Centre for Theoretical Physics, Trieste,
Italy.}\\
{\small \ \ \ }$^{1}${\small \ \ Laboratoire de Physique des Hautes
Energies, Facult\'{e} des Sciences,}\\
{\small \ \ \ \ P.O. Box 1014, Ibn. Battouta, Rabat, Morocco,}\\
{\small \ \ \ \ }$^{2}${\small \ Laboratoire de Physique Th\'{e}orique et
Appliqu\'{e}e (LPTA),}\\
{\small \ \ \ \ Facult\'{e} des Sciences, D\'{e}partement de Physique, B.P.
133, K\'{e}nitra, Morocco}{\tiny .}}
\maketitle

\begin{abstract}
The scope of this work concerns the adaptation of the parallelizability
pp-wave (Ppp-wave) process to $D=10$ type IIB string backgrounds in the
presence of the non-trivial anti-self dual R-R 5-form ${\cal F}$. This is
important in the sense that it gives rise to some unsuspected properties. In
fact, exact solutions of type IIB string backgrounds on Ppp-waves are
discussed. For the $u$-dependence of the dilaton field $\Phi $, we establish
explicitly a correspondence between type IIB supergravity equations of
motion and $2d$-conformal Liouville field theory. We show also that the
corresponding conserved conformal current $T(\Phi)$ coincides exactly with
the trace of the symmetric matrix $\mu _{ij}$ appearing in the quadratic
front factor $F = \mu_{ij}x^{i} x^{j}$ of the Ppp-wave. Furthermore, we
consider the transverse space dependence of the dilaton $\Phi$ and show that
the supergravity equations are easily solved for the linear realization of
the dilaton field. Other remarkable properties related to this case are also
discussed.
\end{abstract}

\hoffset=-1cm \textwidth=11,5cm \vspace*{1cm}

\hoffset=-1cm\textwidth=11,5cm \vspace*{0.5cm}

\newpage

\section{Introduction}

Recently, there has been an increasing interest in string theory
on pp-waves. The motivation comes from the fact that pp-waves
spacetime provides an example of exact string theory background
with all $\alpha ^{^{\prime }}$ corrections vanishing
\cite{ref1,ref2}. The existence of a covariantly constant null
killing vector greatly simplifies the quantization of a string
light cone gauge \cite{ref2, ref3}. In this context, a particular
interest has been devoted to the study of $D=10$ type II string
theory, especially type IIB, on pp-waves, both in NS-NS and R-R
sectors \cite {ref3,ref4,ref5,ref6}. \\ \\
Based on the fact that
all ten dimensional parallelizable solutions (related with the
existence of torsion which makes the manifold flat) of type II or
heterotic supegravities are also $\alpha ^{^{\prime }}$exact
\cite{ref7}, the parallelizable pp-wave backgrounds have been
recently presented in \cite {ref8}. In the same article, the most
general parallelizable pp-wave backgrounds which are non-dilatonic
solutions in the NS-NS sector of type IIA and IIB string theories
are considered.
\\ \\
After that, Figueroa-O'Farrill has classified all the
parallelizable NS-NS backgrounds with the dilaton field turned off
\cite{ref9}, a work which was completed later in type II
supergravity with non trivial dilaton field in \cite{ref10}. The
classification of the simply connected supersymmetric
parallelizable backgrounds of heterotic supergravity was recently
done in \cite{ref11}, where the authors construct all
parallelizable backgrounds of ten dimensional type I supergravity
coupled to supersymmetric Yang Mills.\\ \\

The scope of this work concerns the adaptation of the
parallelizability pp-wave process to type IIB string backgrounds
in $D=10$ in the presence of the anti self dual R-R 5-form ${\cal
F}$. This is a promising topics attracting a lot of attention
recently and showing a big flexibility to open up on new research
ideas. Thus, techniques developed in $2d$-conformal field theories
can be applied to complete several studies about pp-waves and
supergravity solutions. We find among other that in the light cone
dependence of the dilaton field $\Phi $, the ten dimensional
supergravity equations of motion are related to $2d$-conformal
Liouville theory. \\ \\
The outline of this paper is as follows: In
section 2, we give an overview of pp-waves and parallelizability.
The main steps concerning the proof of the theorem relating the
parallelizable pp-waves backgrounds and homogeneous plane waves
are also presented. In section 3, we consider $D=10$ type IIB
supergravity equations of motion, with non trivial R-R 5-form
${\cal F}$, and observe that they are simplified with the
parallelizable pp-waves conditions. Assuming that the R-R 5-form
reads as ${\cal F} = du\wedge \varphi (x^{i})$, with $\varphi
(x^{i})$ is an anti self dual closed 4-form in the transverse
eight dimensional space, we focus to study the $u$-dependence of
the dilaton $\Phi $. This gives rise to a relationship between
type IIB supergravity and conformal Liouville field theory, whose
general solution is explicitly derived. We show also that the
corresponding conserved conformal current $T(\Phi)$, coincides
exactly with the trace of the symmetric matrix $\mu _{ij}$
appearing in the quadratic front factor of the parallelizable
pp-wave $F=\mu _{ij}x^{i}x^{j}$. The other cases corresponding to
the $x^{i}$-dependence of the dilaton $\Phi $, and the transverse
space case are also discussed. Section 4 is devoted to our
conclusion.

\section{PP-Waves and Parallelizability : A Review.}

Being given the importance of supergravity theories which are $\alpha
^{\prime }$-exact solutions, we focus in what follows to review two
essential notions namely the pp-waves and the parallelizability and discuss
the parallelizable pp-waves as well as the homogeneous plane waves. For
other important aspects of these topics, the reader is referred to the
literature \cite{ref8, ref12,ref13,ref14,ref15}.

\subsection{Generalities on PP-waves spacetime}

pp-waves ( plane-fronted waves with parallel ray) are general class of
spacetime admitting a covariantly constant null Killing vector field $v^{\mu
}$

\begin{equation}
\nabla _{\mu }v_{\nu }=0,v^{\mu }v_{\mu }=0
\end{equation}

In general relativity, they form simple solutions to Einstein's equations
with many curious properties. The presence of the covariantly constant null
Killing field implies that these spacetime have vanishing scalar curvature
invariants, much the same as flat space. The most general form of the
pp-waves metrics is given by

\begin{equation}
ds^{2}=-2dudv-F(u,x^{i})du^{2}+2A_{i}(u,x^{j})dudx^{i}+g_{ij}(u,x^{i})dx^{i}dx^{j}.
\end{equation}

To simplify the discussion, it is convenient to work with spacetime for whom
$A_{i}(u,x^{i})=0$. The front factor $F(u,x^{i})$ is shown to satisfy, by
virtue of vacuum Einstein's equations of pure gravity, the transverse
Laplace equation for each $u$ and that the transverse space be Ricci flat.
However, $F(u,x^{i})$ can be considered as an arbitrary function of the
longitudinal coordinate $u$. A useful simplification of the above metrics,
consists in considering pp-wave spacetime with flat transverse part

\begin{equation}
ds^{2}=-2dudv-F(u,x^{i})du^{2}+dx^{i}dx^{i}.
\end{equation}

To restrict much more these classes of pp-waves, one consider the plane
waves spacetime which are those for which the harmonic function $F(u,x^{i})$
is quadratic $F(u,x^{i})=f_{ij}(u)x^{i}x^{j}$. Thus, the plane wave metrics
takes the form

\begin{equation}
ds^{2}=-2dudv-f_{ij}(u)x^{i}x^{j}du^{2}+dx^{i}dx^{i},
\end{equation}

where $f_{ij}(u)$ is any function of the longitudinal coordinate $u$,
symmetric and traceless by virtue of vacuum Einstein's equations.

The homogeneous plane waves further restrict our general pp-wave metric by
taking out $f$'s dependence on $u$. We have

\begin{equation}
ds^{2}=-2dudv-\mu _{ij}x^{i}x^{j}du^{2}+dx^{i}dx^{i},
\end{equation}

where $f_{ij}\equiv \mu _{ij}$ defines a symmetric constant.

A particular example of the homogeneous plane waves is given by the BMN
plane wave metrics \cite{ref3,ref16} for which $f_{ij}=\mu ^{2}\delta _{ij}$.

\subsection{ Parallelizability}

Consider an $n$ dimensional manifold $M$. Due to Cartan-Schouten \cite{ref15}%
, this manifold is said to be parallelizable if there exists a torsion which
``flattens'' the manifold, i.e. makes the Riemann curvature tensor vanish.
Explicitly, we can decompose the connection into a Christoffel piece and a
torsion contribution:

\begin{equation}
\stackrel{\wedge }{\Gamma }_{\mu \nu }^{\lambda }=\Gamma _{\mu \nu
}^{\lambda }+T_{\mu \nu }^{\lambda }
\end{equation}

where $\Gamma _{\mu \nu }^{\lambda }$ is symmetric in $\mu \nu $ indices and
$T_{\mu \nu }^{\lambda }$ (torsion) is anti-symmetric. The curvature $%
\stackrel{\wedge }{R}_{\mu \nu \rho \sigma }$may be decomposed in a similar
way, into a piece which comes only from the Christoffel connection and the
torsional contributions:

\begin{equation}
\stackrel{\wedge }{R}_{\mu \nu \rho \sigma }\equiv R_{\mu \nu \rho \sigma
}+\nabla _{\rho }T_{\mu \nu \sigma }-\nabla _{\sigma }T_{\mu \nu \rho
}+T_{\mu \sigma \lambda }T_{\nu \rho }^{\lambda }-T_{\mu \rho \lambda
}T_{\nu \sigma }^{\lambda }.
\end{equation}

The parallelizability condition is equivalent to set $\stackrel{\wedge }{R}%
_{\mu \nu \rho \sigma }=0$. Furthermore, the manifold is said to be
Ricci-parallelizable if the following Ricci tensor $\stackrel{\wedge }{R}
_{\mu \nu }\equiv R_{\mu \nu }+\nabla _{\lambda }T_{\mu \nu }^{\lambda
}-T_{\mu \sigma \lambda }T_{\nu }^{\sigma \lambda }$ is zero. Requesting the
Ricci-parallelizability yields the vanishing of the symmetric and
antisymmetric pats of $\stackrel{\wedge }{R}_{\mu \nu }$namely

\begin{eqnarray}
R_{\mu \nu }-T_{\mu \sigma \lambda }T_{\nu }^{\sigma \lambda } &=&0,
\nonumber \\
\nabla _{\lambda }T_{\mu \nu }^{\lambda } &=&0.
\end{eqnarray}

In most string theories, where a torsion field arises naturally,
parallelizability leads to important implications for supergravities and
their solutions. The type II NS-NS part of supergravity action is shown to
take the following form

\begin{equation}
{\cal S=}\frac{1}{l_{p}^{8}}\int d^{10}x\sqrt{-g}e^{-2\Phi }\left(
R+4(\nabla _{\mu }\Phi )^{2}-\frac{1}{12}H_{\mu \nu \rho }H^{\mu \nu \rho
}\right).
\end{equation}

The non dilatonic solutions lead to the supergravity equations of motion for
the metric and $B_{\mu \nu }$ field, namely

\begin{eqnarray}
R_{\mu \nu }-\frac{1}{4}H_{\mu \sigma \lambda }H_{\nu }^{\sigma \lambda }
&=&0,  \nonumber \\
\nabla _{\lambda }H_{\mu \nu }^{\lambda } &=&0,
\end{eqnarray}

with $H=dB$. Henceforth, $H_{\mu \sigma \lambda }$ is assumed to be closed $%
dH=0$. Once we define the torsion $T_{\mu \sigma \lambda }$ to be $\frac{1}{2%
}H_{\mu \sigma \lambda }$, the previous supergravity equations are nothing
but the Ricci-parallelizability condition (8).

\subsection{Parallelizable pp-waves and homogeneous plane-waves}

Referring to a theorem presented in \cite{ref8}, in which the authors
demonstrate that parallelizable pp-waves backgrounds are necessarily
homogeneous plane waves and that a large class of homogeneous plane waves
are parallelizable stating the necessary conditions. We remind here bellow
the main steps of this proof. The starting point is to consider the most
general ten dimensional pp-wave geometry whose metric is given in (2) where
the index $i,j$ stands for the transverse coordinates $i,j=1,2,...,8.$ The
functions $F,A_{i},g_{ij} $ are chosen to satisfy supergravity equations of
motion. Next, the most general NS-NS $H$ field compatible with the
covariantly constant Killing vector $v^{\mu }$ is given by

\begin{equation}
H_{uij}=h_{ij}(u,x^{k}),
\end{equation}

and all the remaining components zero. Therefore, by imposing the
Ricci-parallelizability conditions, it's pointed out that the only
nonvanishing component of the Ricci curvature is $R_{uu}$. Performing some
transformations based on the parallelizability condition, the authors
present the most general parallelizable pp-waves

\begin{eqnarray}
ds^{2} &=&-2dudv-\mu _{ij}x^{i}x^{j}du^{2}+dx^{i}dx^{i} \\
H_{uij} &=&h_{ij}=constant.
\end{eqnarray}

with

\begin{equation}
\mu _{ij}=\frac{1}{4}h_{ik}h_{jk},
\end{equation}

which is of the form of an homogeneous plane wave, a fact which complete the
proof. As claimed also in \cite{ref8}, for the inverse case, not all
homogeneous plane-wave geometries are parallelizable. The exception is made
only for homogeneous plane wave for which $\mu _{ij}$ has doubly degenerate
eigenvalues

\begin{equation}
\mu
_{ij}=diag(a_{1}^{2},a_{1}^{2},a_{2}^{2},a_{2}^{2},a_{3}^{2},a_{3}^{2},a_{4}^{2},a_{4}^{2})
\end{equation}

where the four real numbers $a_{i}$, determining completely the
parallelizable pp-wave, are expressed in terms of the antisymmetric $8\times
8$ matrix $h_{ij}$ whose only non-zero components are

\begin{equation}
h_{12}=2a_{1},h_{34}=2a_{2},h_{56}=2a_{3},h_{78}=2a_{4}.
\end{equation}

\section{Type IIB string backgrounds equations on parallelizable PP-waves}

\subsection{ Parallelizable pp-waves backgrounds:}

The starting point is the lowest order effective action of type IIB
supergravity in $10$-dimensions with non-trivial R-R 5-form ${\cal F}$ \cite
{ref5}.

\begin{equation}
{\cal S}_{eff}{\cal =}\int d^{10}x\sqrt{-g}e^{-2\Phi }\left( R+4(\nabla
_{\mu }\Phi )^{2}-\frac{1}{12}H_{\mu \nu \rho }H^{\mu \nu \rho }-\frac{%
e^{2\Phi }}{4.5!}{\cal F}_{\mu \nu \rho \sigma \tau }{\cal F}^{\mu \nu \rho
\sigma \tau }\right) ,
\end{equation}

in the presence of a metric $g$, a non trivial dilaton $\Phi $, an
anti-symmetric NS-NS $B$ field with field strength $H=dB$. Setting $%
C^{(0)}=0=C^{(2)}$ for the remaining R-R fields, we may write ${\cal %
F=\partial }C^{(4)}$ where $C^{(4)}$ is the corresponding R-R 4-form
appropriately normalized. In $10$ dimensions, the self duality of ${\cal F}$
namely $^{*}{\cal F=F}$ does not admit a natural derivation from a covariant
action principle and is imposed as an additional constraint at the level of
the derived supergravity equations of motion in the $\sigma $-model frame

\begin{eqnarray}
R_{\mu \nu } &=&-2\nabla _{\mu }\nabla _{\nu }\Phi +\frac{1}{4}H_{\mu \rho
\sigma }H_{\nu }^{.\rho \sigma }+\frac{e^{2\Phi }}{4.4!}\left( {\cal F}_{\mu
\kappa \lambda \rho \sigma }{\cal F}_{\nu }^{.\kappa \lambda \rho \sigma }-%
\frac{1}{10}g_{\mu \nu }{\cal F}_{\kappa \lambda \rho \sigma \tau }{\cal F}%
^{\kappa \lambda \rho \sigma \tau }\right) ,  \nonumber \\
0 &=&\nabla _{\mu }\nabla ^{\mu }\Phi -2\left( \nabla _{\mu }\Phi \right)
\left( \nabla ^{\mu }\Phi \right) +\frac{1}{12}H_{\mu \nu \rho }H^{\mu \nu
\rho },  \nonumber \\
0 &=&\nabla _{\mu }(e^{-2\Phi }H^{\mu \nu \rho }),  \nonumber \\
0 &=&\nabla _{\mu }{\cal F}^{\mu \nu \kappa \lambda \rho },
\end{eqnarray}

where $\nabla _{\mu }$ is the covariant derivative with respect to the
Levi-Civita connection $\Gamma _{\mu \rho }^{\nu }$. The Greek indices
stands for the values $0,1,...,9.$ The field strength $H_{\mu \nu \rho }$ of
the NS-NS two-form $B_{\mu \nu }$ is assumed to be a closed 3-form, $dH=0.$\

Next, we focus to adapt the parallelizability framework reviewed in section
2, to the previous supergravity equations a fact which give rise to some
unsuspected properties. Indeed, requesting the Ricci parallelizability
conditions

\begin{eqnarray}
R_{\mu \nu } &=&\frac{1}{4}H_{\mu \rho \sigma }H_{\nu }^{.\rho \sigma },
\nonumber \\
0 &=&\nabla _{\mu }H^{\mu \nu \rho },
\end{eqnarray}

equations (18) reduce then to

\begin{eqnarray}
0 &=&-2\nabla _{\mu }\nabla _{\nu }\Phi +\frac{e^{2\Phi }}{4.4!}\left( {\cal %
F}_{\mu \kappa \lambda \rho \sigma }{\cal F}_{\nu }^{.\kappa \lambda \rho
\sigma }-\frac{1}{10}g_{\mu \nu }{\cal F}_{\kappa \lambda \rho \sigma \tau }%
{\cal F}^{\kappa \lambda \rho \sigma \tau }\right) ,  \nonumber \\
0 &=&\nabla _{\mu }\nabla ^{\mu }\Phi -2\left( \nabla _{\mu }\Phi \right)
\left( \nabla ^{\mu }\Phi \right) +\frac{1}{12}H_{\mu \nu \rho }H^{\mu \nu
\rho },  \nonumber \\
0 &=&H^{\mu \nu \rho }\nabla _{\mu }e^{-2\Phi },  \nonumber \\
0 &=&\nabla _{\mu }{\cal F}^{\mu \nu \kappa \lambda \rho },
\end{eqnarray}

Note by the way, that once ${\cal F=}0,$ the above equations reduce to

\begin{eqnarray}
\triangledown _{\mu }\triangledown _{\nu }\Phi &=&0,  \nonumber \\
\triangledown _{\mu }\Phi \triangledown ^{\mu }\Phi &=&\frac{1}{24}H_{\mu
\nu \rho }H^{\mu \nu \rho },  \nonumber \\
H^{\mu \nu \rho }\triangledown _{\mu }e^{-2\Phi } &=&0.
\end{eqnarray}

Therefore, as mentioned in \cite{ref10}, it follows from the above equations
that non-dilatonic background solutions are viable if $H_{\mu \nu \rho
}H^{\mu \nu \rho }=0$, which is equivalent to set the scalar curvature $R$
to be zero. This corresponds to the parallelizable pp-waves discussed in
\cite{ref8, ref9}. However, for $H_{\mu \nu \rho }H^{\mu \nu \rho }\neq 0,$
it's also shown that the set of equations (21) can be solved easily with
linear dilaton field $\Phi $ in the direction transverse to $H_{\mu \nu \rho
}.$

Since our interest focuses on the ${\cal F}\neq 0$ case, we use the most
general parallelizable pp-wave \cite{ref8}.

\begin{eqnarray}
ds^{2} &=&-2dudv-\mu _{ij}x^{i}x^{j}du^{2}+dx^{i}dx^{i} \\
H_{uij} &=&h_{ij}=constant.
\end{eqnarray}

in order to simplify the string background equations (20) where the constant
$\mu _{ij}$ and $h_{ij}$ are given by (14-16). As we will show later, the
type IIB supergravity solutions in the presence of the R-R 5-form is
associated to a non linear conformal Liouville realization of the dilaton
field. This is an unusual property which goes beyond the special NS-NS case $%
{\cal F}=0$ and which provides then a possibility to connect conformal
models with string background solutions.

Thereafter, as we are interested to simplify more the string background
equations, we assume for the moment that the dilaton $\Phi $ is arbitrary
and consider the following realization for the 5-form \cite{ref5},

\begin{equation}
{\cal F=}du\wedge \varphi (x^{i})
\end{equation}

where $\varphi (x^{i})$ is an anti-self dual closed 4-form in the transverse
eight dimensional space with

\begin{eqnarray}
\varphi (x^{i}) &=&-^{*}\varphi (x^{i})  \nonumber \\
d\varphi (x^{i}) &=&0
\end{eqnarray}

such that ${\cal F=}^{*}{\cal F}$. Thus, taking the $uu$-component of the
first equation of (20), we have to consider different cases,

A) $\Phi =\Phi (u)$ one obtains

\begin{equation}
0=-2\nabla _{u}\nabla _{u}\Phi +\frac{e^{2\Phi }}{4.4!}\varphi
_{ijkl}\varphi ^{ijkl}
\end{equation}

This case gives rise to non standard results as we will explicitly discuss
in the next subsection. Later on we will use the following notation $\varphi
_{ijkl}\varphi ^{ijkl}$ $\equiv \varphi ^{2}$ for the square of the 4-form.

B) $\Phi =\Phi (x^{i})$ one obtains

\begin{equation}
0=\varphi _{ijkl}\varphi ^{ijkl}
\end{equation}

Before any comments about this equation, lets remind a different formula
found by the authors of \cite{ref5} in the same context but without
requesting the parallelizability

\begin{equation}
\nabla _{i}\left( e^{-2\Phi }\nabla ^{i}F\right) =\frac{1}{2.4!}\varphi
_{ijkl}\varphi ^{ijkl}.
\end{equation}

This equation, obtained by taking the $uu$-component of the non
parallelizable supergravity equation of motion, gives a relation between the
front factor $F$ and the anti-self dual closed 4-forme $\varphi$. Thus, it
should undergo an important change while requiring the parallelizability.
Indeed, once the parallelizability condition is applied to string background
equations of motion (18) for $\Phi =\Phi (x^{i}),$ the previous ($\varphi $,$%
F)$-coupling is not anymore allowed, since the derivative term associated to
the front $F$ although non zero decouples from $\varphi _{ijkl}\varphi
^{ijkl}$ which becomes zero as given in (27).

For this case, we point out that the remaining string background equations

\begin{eqnarray}
0 &=&\nabla _{i}\nabla ^{i}\Phi -2\left( \nabla _{i}\Phi \right) \left(
\nabla ^{i}\Phi \right) +\frac{1}{12}h^{2},  \nonumber \\
0 &=&H^{uij}\nabla _{i}e^{-2\Phi },
\end{eqnarray}

with $h^{2}\equiv H_{uij}H^{uij}$; can be solved by a linear realization of
the dilaton field $\Phi (x^{i})$ which provides also a solution for the
first equation of (20) once $\varphi ^{2}=0$.

However, in the transverse space, the equations of motion (20) reduce to

\begin{eqnarray}
0 &=&-2\nabla _{i}\nabla _{j}\Phi,  \nonumber \\
0 &=&\nabla _{i}\nabla ^{i}\Phi -2\left( \nabla _{i}\Phi \right) \left(
\nabla ^{i}\Phi \right) ,
\end{eqnarray}

since the only nonvanishing components of $H$-field are $H_{uij}=h_{ij}$ and
therefore $H_{ijk}=0$. The first equation shows that the dilaton field $\Phi
$ should be linear in the transverse coordinates and based on this fact the
second one corresponds to the vanishing of the dilaton gradient $\nabla
_{i}\Phi $. As a results the dilaton is constrained to be a constant in the
transverse coordinates.

\subsection{Parallelizable PP waves and conformal Liouville field Theory}

We focus in this subsection to give more details about the derived
supergravity equation of motion presented in the first case (26) since it
gives rise to important non standard results. The idea behind our analysis
consists in complexifying the above equation, a fact which makes the link
with $2d$-conformal symmetry possible. Indeed, consider the differential
equation (26), for some reasons that we will explain later, let's suppose
that one can temporarily introduce by hand the derivation with respect to
the $v$-longitudinal direction so that the equation becomes\footnote{%
It's important to note at this level that the $v$-longitudinal direction
ignored in the standard supergravity backgrounds formulation is now
resuscitated to build a $2d$-conformal filed theory framework.}

\begin{equation}
0=-2(\nabla _{u}^{2}+\nabla _{v}^{2})\Phi +\frac{e^{2\Phi }}{4.4!}\varphi
_{ijkl}\varphi ^{ijkl}
\end{equation}

This way to write the things made us recall the complex notation and $2d$%
-conformal invariant equations. Indeed, adopting this analogy for which we
put,

\begin{equation}
\nabla _{u}^{2}+\nabla _{v}^{2}\equiv \partial \overline{\partial },
\end{equation}

with $\partial =\frac{\partial }{\partial z}\equiv \nabla
_{u}+i\nabla _{v}$ and $\overline{\partial }=\frac{\partial
}{\partial \overline{z}}\equiv \nabla _{u}-i\nabla _{v}.$ This
drives us to rewrite the supergravity equation of motion (26) as a
bidimensional conformal Liouville like equation, namely

\begin{equation}
\partial \overline{\partial }\Phi -{\chi }(\varphi )\exp (2\Phi )=0,
\end{equation}

or equivalently

\begin{equation}
\beta \partial \overline{\partial }\Phi -{\chi }(\varphi )\exp (2\beta \Phi
)=0,
\end{equation}

modulo the following scaling $\Phi \rightarrow \beta \Phi$, where $\beta $
is a real constant parameter and where the coefficient ${\chi }(\varphi )$
stands for the constant ${\chi }(\varphi )\equiv \frac{\varphi ^{2}}{4!.8}$.
By virtue of our knowledge on $2d$-CFT and integrable models \cite
{ref18,ref19,ref20}, this equation is shown to be solvable as we will show
later.

The fact to add the derivation with respect to the $v$-longitudinal
coordinate is only a formal trick of calculation that is going to be ignored
thereafter. Note also that we have omitted in the above complexification the
fact that $\nabla _{u}$designates really the covariant derivatives. This
omission doesn't influence our analysis and the comparison with the
conformal Liouville equation imposes itself by nature.

Now before discussing the conformal properties and the solution of this new
equation (34), for the meantime we are going to recall some known properties
of the standard Liouville equation \cite{ref21,ref22}.

The local equation of motion of the two dimensional Liouville field $\phi (z,%
\overline{z})$ is given by

\begin{equation}
\beta \partial \overline{\partial }\phi -\exp (2\beta \phi )=0,
\end{equation}

where $\beta $ is a real coupling constant. $2d$ Liouville equation was the
subject of several studies in the literature and in different contexts \cite
{ref23,ref24}. It is a non linear differential equation that has the
particularity to be conformally invariant. Because of the rich structure of
bidimensional conformal models, the conformal symmetry represents in this
setting a guarantee of the solvability (integrability) and therefore assures
a means to overcome the non linearity.

To study the integrability of (35), different techniques including the Lax
method were developed. Here, we content ourselves to recall that the
explicit solution of the nonlinear Liouville equation can be written as \cite
{ref22}

\begin{equation}
\exp (2\beta \phi )=k\frac{f^{\prime }(z).\overline{f}^{\prime }(\overline{z}%
)}{\left( 1-f(z).\overline{f}(\overline{z})\right) ^{2}},
\end{equation}

where $f(z)$ and $\overline{f}(\overline{z})$ are arbitrary analytic and
anti-analytic functions, $f^{\prime }(z)=\partial f(z)$ and $\overline{f}%
^{\prime }(\overline{z})=\overline{\partial (}\overline{f}(\overline{z}))$
and $k$ is an arbitrary constant for the moment. As it is well known, the
integrability of the Liouville equation is due to conformal symmetry
generated by the following classical energy momentum tensor

\begin{equation}
T(\varphi )=(\partial \varphi )^{2}-\frac{1}{\beta }(\partial ^{2}\varphi ).
\end{equation}

The conservation law of this conformal current, follows immediately by using
the equation of motion as shown here below

\begin{eqnarray}
\overline{\partial }T(\varphi ) &=&2(\partial \overline{\partial }\varphi
)(\partial \varphi )-\frac{1}{\beta }\partial (\partial \overline{\partial }%
\varphi )  \nonumber \\
&=&-\frac{1}{\beta }\left( \partial -2\beta \partial \varphi \right)
\partial \overline{\partial }\varphi  \nonumber \\
&=&0.
\end{eqnarray}

Now having given the necessary ingredients of the conformal Liouville
equation, we pass now to study our equation (34). The non-trivial behavior
that ensues from our analysis is based on the fact that the striking
resemblance with conformal Liouville equation is going to allow us to adapt
conformal symmetry and integrable models backgrounds to the present context
of parallelizable pp-waves.

Another point to evoke concerns the limit to consider in order to recover
the standard formalism where the supergravity equations are expressed
according to the covariant derivative \footnote{%
This is because the $v$-direction is ignored in the standard supergravity
computations.} $\nabla_{u}$ and not of the complex ones $\partial $ and $%
\overline{\partial }$. Such a limit is given simply by performing the
following transformations at the level of the derived complex formulas

\begin{eqnarray}
z,\overline{z} &\equiv &u  \nonumber \\
\partial .\overline{\partial } &\equiv &\nabla _{u}^{2}  \nonumber \\
\partial f(z) &\equiv &\nabla _{u}f(u)\equiv \overline{\partial (}\overline{f%
}(\overline{z}))
\end{eqnarray}

Now, consider the derived $2d$ conformal Liouville equation of motion (34).
This equation describes the results of the parallelizable pp-waves
constraint on the string backgrounds given by the supergravity effective
action (17). The direct contact with the standard Liouville equation (35)
consists in setting for instance as an ansatz

\begin{equation}
{\chi }(\varphi )\equiv \frac{\varphi ^{2}}{4!.8}=1.
\end{equation}

For this simple choice, our derived Liouville equation (34) is
solved and the solution is given by (36) for any arbitrary
analytic and anti-analytic functions $f(z)$ and $\bar{f}(\bar{z})$
respectively. One should notice at this level the importance of
the complex formulation since it gives rise to several solutions
related the arbitrary character of the functions $f(z)$ and
${\bar{f}}(\bar{z})$. As an example, consider the linear
realization $f(z)=az$ and ${\bar{f}}(\bar{ z})=b\bar{z}$. One
easily check that this linear choice corresponds to a solution of
the Liouville equation (34) and in the same way fixes the constant
$k$ to take the value $k=1$. Such a richness in the solutions is a
natural feature of $2d$-conformal symmetry, a property which is no
longer guaranteed if the conformal symmetry is lost.
\\ \\
Now, as we need to recover our derived string backgrounds equations (26),
one have to perform the limit equations (39) given above. A remarkable fact
here, is that the introduction of these limit equations break automatically
the conformal symmetry and the previous analysis must be controlled with
prudence. The loss of the conformal symmetry, can be traced to the fact that
the contribution of the $v$-longitudinal coordinate is annulled. Indeed, the
conformal splitting in two different, analytic and anti-analytic, sectors is
not more valid since we will have only one sector described by the
contribution of the $u$-coordinate.
\\ \\
To illustrate these ideas for the simple case ${\chi }(\varphi )=1$, let's
consider the above $2d$-conformal approach in the limit (39). The Liouville
equation of motion (34) is now reduced to the original string backgrounds
equation (26) (the $u$-Liouville equation) whose formal solution is assumed
for the moment to derive from the complex one (36), once the limit (39) are
performed, we set

\begin{equation}
\exp (2\beta \Phi )=k\left( \frac{\nabla _{u}f(u)}{1-f^{2}(u)}\right) ^{2}.
\end{equation}

As discussed above, the fact to ignore the contribution of the $v$%
-coordinate will certainly induces some constraint on the functions $f(u)$
that are not more arbitrary. Note that we preserved the shape (41) of the
solutions to assure the compatibility with the above approach. Also, this
equation shows how the dilaton field $\Phi $ is explicitly expressed in
terms of the $\nabla _{u}$-derivatives of some constrained $u$-dependent
functions $f(u)$ that we will try to discuss.
\\ \\
Thereafter, we are interested to derive explicitly the functions $f(u)$ for
which (41) defines a solution of (26). Simple formal computations show that
the $u$-Liouville equation (26) has in fact two explicit solutions for the
dilaton $\Phi(u)$ namely

\begin{equation}
\Phi _{\pm }(u)=\log \left[ \pm a\sqrt{-1+\tanh ^{2}(au+b)}\right]
\end{equation}

or equivalently

\begin{equation}
\Phi _{\pm }(u)=\log [\pm ia\sec h(au+b)]
\end{equation}

where $a$ and $b$ are two positive constant numbers with $\sec h(\alpha )=
\frac{1}{\cosh (\alpha )}$. Actually, the two explicit solutions of the
equation of motion (26), can be extracted from our proposed solution (41).
Indeed, performing straightforward by lengthy computations show that the two
derived solutions are described by two doubly degenerate expressions of the
the function $f(u)$ namely

\begin{equation}
f_{\pm }(u){=}\left[ -1+\frac{2}{1+\exp [\pm \frac{4i\arctan [\tanh (\alpha
u+\beta )]}{\sqrt{k}}+\gamma ]}\right] ^{-1}{,}
\end{equation}

with $\alpha =\frac{1}{2}a$, $\beta =\frac{1}{2}b$ and $\gamma $ are
constant numbers. We note by the way that the $u$-Liouville equation (26)
possesses also a simple solution namely \footnote{%
We thank M. Blau for fruitful discussions and suggestions.}
\begin{equation}
\Phi (u)=-\log [\cos (au)],
\end{equation}
such a solution is satisfied for the constant $a$ such that ${\chi }(\varphi
)\equiv \frac{\varphi ^{2}}{4!.8}=a^{2}.$ Its also important to search, at
the level of general formula (41), for the associated function $f(u)$. In
fact the same analysis drives us to write for the simple case $a=1$%
\begin{equation}
f{(u)=}\frac{K_{1}}{\sin (u)-K_{2}}-K_{2},
\end{equation}
where $K_{1}$ and $K_{2}$ are two constants related as follows $%
K_{1}=1-K_{2}^{2}$ . One can signal at this level the importance of the
conformal approach described above and from which we have been able to
derive the expression of the general solution (41). The previous results
show clearly how this solution is more general as it incorporates most of
the solutions discussed. This property, gives in some sense a possibility to
classify all the existing solutions of the $u$-Liouville string backgrounds
equation of motion (26) and describes in the same time an inverse problem
based on the fact that finding $f(u)$ such that (26) is true is a guarantee
of the integrability.
\\ \\
Next, using the same analysis described before, the conserved conformal
current leads in the limit case to the following equation
\begin{equation}
T(\Phi )=(\nabla _{u}\Phi )^{2}-\frac{1}{\beta }(\nabla _{u}^{2}\Phi ),
\end{equation}

To understand the meaning of this conserved current in the string background
supergravity equations, one recall that the remaining equations (20) behave
with respect to the $u$-dependence of the dilaton field as

\begin{equation}
0=\nabla _{u}^{2}\Phi -2\left( \nabla _{u}\Phi \right) ^{2}+\frac{1}{12}%
h^{2}.
\end{equation}
Identifying (47) and (48), one may extract then the following important
relations
\begin{equation}
T(\Phi )\equiv \mu _{ii}=\frac{1}{4}h^{2}=2\sum_{i=1}^{4} a_{i}^{2}.
\end{equation}
This is an unusual form of the stress energy momentum tensor whose
conservation law follows naturally by using the Liouville equation of motion
since

\begin{equation}
\mu _{ii}=\left( \nabla _{u}\Phi \right) ^{2}-\frac{1}{2}\nabla _{u}^{2}\Phi
,
\end{equation}
is a constant ( with respect to the $u$-direction). We have
\begin{eqnarray}
\overline{\partial }T(\Phi ) &\equiv &\nabla _{u}T(\Phi )  \nonumber \\
&=&\nabla _{u}\mu _{ii}  \nonumber \\
&=&0
\end{eqnarray}
Remark also that the previous formulas impose for the $\beta $-parameter to
take the value $\beta =2.$

Now having discussed a special case associated to the ansatz (40), we intend
thereafter to present the general solution of the conformal Liouville
equation $\beta \partial \overline{\partial }\Phi -{\chi }(\varphi )\exp
(2\beta \Phi )=0,$ for any arbitrary value of the constant ${\chi }(\varphi
) $ . As we can easily check, the conformal conserved current associated to
this Liouville equation does not depend on ${\chi }(\varphi )$ and conserve
then the same shape (47). Now, let's assume that the solution can be written
as follows

\begin{equation}
\exp (2\beta \Phi )=\eta \frac{f^{\prime }(z).\overline{f}^{\prime }(%
\overline{z})}{\left( 1-f(z).\overline{f}(\overline{z})\right) ^{2}},
\end{equation}
where we have to determine later the constant $\eta $ in terms of ${\chi }%
(\varphi )$. Indeed, straightforward but lengthy calculations show that the
quantities ${\chi }(\varphi )$ and $\eta $ are forced to satisfy the
following relation

\begin{equation}
\eta .{\chi }(\varphi )=2,
\end{equation}
and then the final solution is given by
\begin{equation}
\exp (2\beta \Phi )=\frac{2}{{\chi }(\varphi )}.\left( \frac{\nabla _{u}f(u)%
}{1-f^{2}(u)}\right) ^{2},
\end{equation}
once the limit equations are performed. Setting ${\chi }(\varphi )=1$ one
recover the special case discussed before and also fixes the constant $k$ to
be $k=2$.

As a final point, from the third equation of (20) namely

\begin{equation}
0=H^{uij}\nabla _{u}e^{-2\Phi }
\end{equation}

we learn that the dilaton gradient introduced in the $u$-direction should,
by virtue of (26), satisfy a non linear conformal Liouville equation. This
results goes beyond the one presented in \cite{ref10} in which a linear
solution for the dilaton field in the u-direction is shown to deal with the
absence of the Ramond- Ramond 5-form.

\section{Conclusion}

In this work, we present an adaptation of the parallelizability process to
type IIB string backgrounds in $D=10$ dimensions. Essentially, we show that
the two dimensional conformal Liouville theory may be involved in this sense
in a natural way. This is an unsuspected property based on the assumption
that the R-R $5$-form ${\cal F}$ is nonvanishing a fact which leads to exact
solutions of type IIB string backgrounds on parallelizable pp-wave.
\\ \\
The link with conformal field theories is made through the
bidimensional Liouville model. It is a model based on a nonlinear
differential equation and that possesses the particularity to be
conformally invariant and integrable. This conformal Liouville
equation that appears naturally, at the level of the string
background solutions, can be traced to the fact that the dilaton
field depends on the longitudinal coordinate $u$. This hypothesis
is also reinforced with the fact that the R-R $5$-form ${\cal F}$
is non trivial $\left( \varphi _{ijkl}\varphi ^{ijkl}\neq 0\right)
$ since at the level of the equation (31), the Liouville potential
$exp\left( 2\Phi \right) $ is coupled to $\varphi _{ijkl}\varphi
^{ijkl}.$
\\ \\
We show also that the corresponding conserved conformal current $T(\Phi)$
coincides exactly with the trace of the constant symmetric matrix $\mu _{ij}$
appearing in the quadratic front factor of the parallelizable pp-wave namely
$F=\mu _{ij}x^{i}x^{j}$. Furthermore, we consider the transverse space
dependence of the dilaton $\Phi $ and show that the supergravity equations
are easily solved with the linear dilaton field. Other remarkable features
are also discussed.
\\ \\
Finally, it might be of particular interest to study string theory on the
background developed in this work. Also, in the same context, it would be
nice to see if $2d$-sigma model is solvable at least for some special cases
and to look for the supersymmetric behavior of the present analysis\footnote{%
We acknowledge M.M.Seikh-Jabbari for suggesting these important points}.
\newline
\newline
\newline
\newline

{\bf Acknowledgments}\newline

The authors would like to thank the Abdus Salam International Centre for
Theoretical Physics, Trieste-Italy, and the considerable help of the High
Energy Section where this work was done. This work was done within the
framework of the Associateship scheme of the Abdus Salam ICTP (Summer 2003).
M.B.S also wishes to thank M.Blau and M.M.Seikh Jabbari for useful
discussions and important suggestions. 

\end{document}